\documentclass[12pt]{iopart}
\usepackage{iopams}  
\usepackage{graphicx,color}% Include figure~ files
\usepackage{epsfig}
\usepackage{mathrsfs}
\usepackage{bm}% bold math

 \newcommand{\bra}[1]{\langle{#1} |}
 \newcommand{\ket}[1]{|{#1}\rangle  }
 
 \newcommand{\ketbra}[2]{\vert {#1} \rangle \!\langle{#2}\vert}

 \providecommand{\openone}{\leavevmode\hbox{\small1\kern-3.8pt\normalsize1}}

\begin{document}

\title{Hidden entanglement in the presence of random telegraph dephasing noise}

\author{Antonio D'Arrigo$^{1,2}$, Rosario Lo Franco$^{3,4}$, Giuliano Benenti$^{5,6}$, Elisabetta Paladino$^{1,2,4}$, 
and Giuseppe Falci$^{1,2,4}$}
\address{$^1$Dipartimento di Fisica e Astronomia,
Universit\`a degli Studi Catania, Via Santa Sofia 64, 95123 Catania, Italy,\\
$^2$CNR-IMM-UOS Catania (Universit\`a), Consiglio Nazionale delle Ricerche,
Via Santa Sofia 64, 95123 Catania, Italy,\\
$^3$CNISM and Dipartimento di Fisica, Universit\`a di Palermo, 
via Archirafi 36, 90123 Palermo, Italy,\\
$^4$Centro Siciliano di Fisica Nucleare e di Struttura della 
Materia (CSFNSM), Viale S. Sofia 64, 95123 Catania, Italy,\\
$^5$CNISM and Center for Nonlinear and Complex Systems,
Universit\`a degli Studi dell'Insubria, Via Valleggio 11, 22100 Como, Italy,\\
$^6$Istituto Nazionale di Fisica Nucleare, Sezione di Milano,
via Celoria 16, 20133 Milano, Italy
}

\begin{abstract}
Entanglement dynamics of two noninteracting qubits, locally affected by random telegraph noise at pure dephasing, 
exhibits revivals.   
These revivals are not due to the action of any nonlocal operation, thus their occurrence may appear paradoxical since
entanglement is by definition a nonlocal resource. We show that a simple explanation of this phenomenon
may be provided by using the (recently introduced) concept of \emph{hidden} entanglement, which signals the presence of entanglement 
that may be recovered with the only help of local operations.
\end{abstract}

\maketitle

\section{Introduction}
Entanglement is one of the most peculiar features of quantum mechanics and it also
plays the role of a fundamental resource in many applications of quantum information~\cite{nielsen-chuang, benenti-casati-strini, PlenioReview, 
HorodeckiReview}.
On the other hand, entangled systems unavoidably interact with their environments causing decoherence and
a loss of entanglement. Since entanglement is by definition a nonlocal 
resource, one expects that any attempt 
to restore it must involve the use of nonlocal operations. 

We consider physical situations where two subsystems, for example 
two qubits, are prepared in an entangled state and subsequently 
decoupled~\cite{paladinoNJP,darrigoNJP}. Due to the interaction with their 
local environment, entanglement dynamics may exhibit a
non-monotonous behaviour, with the occurrence of revivals alternating to dark periods~\cite{bellomo2007PRL,lopez2008PRL,revivalstochastic,rossini,lofrancorandom,lofrancoreview}.
In some cases, this phenomenon is due to the fact that entanglement
is transferred to  quantum environments, and then 
back-transferred to the system~\cite{bellomo2007PRL,lopez2008PRL}.
In other cases, the environment can be modelled as a 
classical system~\cite{revivalstochastic,lofrancorandom,lofranco2012}
and no entanglement between the system and the environment is established at any 
time. In the latter cases, the occurrence of entanglement revivals
may appear paradoxical, since the effect of the noise is analogous to a local operation on a subsystem. A first interpretation of this phenomenon has been given in terms of correlations present in a classical-quantum state of environments and qubits \cite{lofrancorandom}. 
In a recent work~\cite{darrigo2012}, we have proposed to solve
the apparent paradox by introducing the concept of \emph{hidden}
entanglement (HE), which measures the amount of entanglement 
that may be recovered without the help of any nonlocal operation.
The definition of HE is based on the quantum trajectory description
of the system dynamics, that allows to point out the presence of entanglement in the system
even if the density operator formalism does not reveal it:
this entanglement is thus not accessible (hidden) due to the lack of 
classical information~\cite{darrigo2012}.

Relevant examples of situations where the environment can be modelled as a classical
system may be found in solid-state implementations of qubits.
For example, in superconducting nanocircuits one of the most relevant sources of decoherence~\cite{FalciPrl05,solidstate,nakamura02} 
are fluctuating background charges localized in the insulating materials surrounding superconducting islands~\cite{zorin96}. 
Each impurity produces a bistable fluctuation of the island polarization. 
The collective effect of an ensemble of these random telegraph processes, with a proper distribution of
switching rates, gives rise to $1/f$-noise~\cite{weissman} routinely 
observed nanodevices~\cite{nakamura02,zorin96,kafanov08}.

In this paper we exploit the concept of hidden entanglement to explain the occurrence of entanglement revivals in a 
simple system. In particular, we consider two noninteracting
qubits, one of them affected by a random telegraph noise 
at pure dephasing~\cite{paladino02,paladino03,galperin06}. 
The paper is organized as follows. In section \ref{sec:model}, we introduce
the Hamiltonian model.
In section \ref{sec:entanglement-dynamics} we discuss the
entanglement dynamics, showing that the revivals of entanglement 
are due to the presence of hidden entanglement. In section \ref{conclusions} we summarize obtained results and 
present some final comments.

\section{Model}
\label{sec:model}
We consider two noninteracting qubits $A$ and $B$, initially prepared
in a pure maximally entangled state 
$\rho(0)=\ket{\varphi}\bra{\varphi}$, evolving
according to the Hamiltonian ($\hbar=1$)
\begin{eqnarray}
&&\hspace{0.0cm}{\cal H}\,=\,{\cal H}_0\,+\,\delta {\cal H},\nonumber\\
&&\hspace{0.0cm}{\cal H}_0=\,-\frac{\Omega_A}{2}\sigma_{z_A}-\frac{\Omega_B}{2}\sigma_{z_B},\qquad
\delta{\cal H}=-\frac{\xi(t)}{2}\sigma_{z_A}, \label{eq:Hamiltonian1}
\end{eqnarray} 
where $\sigma_{z_A}=\sigma_{z}\otimes \openone$, 
$\sigma_{z_B}=\openone\otimes\sigma_{z}$ and $\delta \cal H$ represents 
a random telegraph  (RT)  process $\xi(t) 
\in \{0,v\}$~\cite{Papoulis} acting on  qubit $A$. 
The RT process induces a random switching of qubit $A$ frequency between
$\Omega_A/(2\pi)$ and $(\Omega_A+v)/(2\pi)$, with an overall  
switching rate $\gamma$ (without loss of generality, we assume $v>0$). 
We consider a symmetric RT process where the 
transition rates between the two  states  are equal, that is
$\gamma_{0 \rightarrow v}=\gamma_{v \rightarrow 0}=\gamma/2$.

Our first aim is to find the system density matrix at any time $\rho(t)$.
The two qubits independently evolve under the Hamiltonian (\ref{eq:Hamiltonian1}): 
qubit $B$ freely evolves whereas qubit $A$ 
displays a pure dephasing dynamics due to the effect of the stochastic process $\xi(t)$.
The dynamics of single qubit subject to RT noise at pure dephasing has been solved 
in~\cite{paladino02,paladino03,galperin06}. 
A possible way to obtain $\rho(t)$ is to solve a \textit{stochastic}  
Schr\"odinger equation which gives the following formal expression for $\rho(t)$ 
\begin{equation}
\rho(t)=\int {\cal D}[\xi(t)] P[\xi(t)]\,\rho_\xi(t) \, ,
\label{eq:evolved_qubit_densityoperator}
\end{equation}
where $\rho_\xi(t)=\ketbra{\varphi_{\xi}(t)}{\varphi_{\xi}(t)}$ with 
$\ket{\varphi_{\xi}(t)}=e^{\frac{i}{2}\int_0^t\xi(t')dt'\sigma_{z_A}}e^{-i{\cal H}_0 t}\ket{\varphi}$,
and the probability of the realization $\xi(t)$ can be written as  
\begin{equation}
P[\xi(t)]=\lim_{m\to \infty} \eta_{m+1}(\xi_m,t_m; \ldots;\xi_1,t_1;\xi_0,t_0),
\label{eq:probability_evolved_qubit_state}
\end{equation}
where $\eta_{m+1}$  is a ($m+1$) joint probability for the sampled
$\xi(t)$ at regular intervals 
$\Delta t=(t-t_0)/m$, $t_k=t_0+k\Delta t$, 
$\xi_k\equiv\xi(t_k)$ ($k=0,\ldots,m$)~\cite{PaladinoPhScr09}.
Since the qubits evolve independently, the above procedure leads to a simple form depending on the
single qubit coherences.
In the computational basis $\{\ket{00},\ket{01},\ket{10},\ket{11}\}$, where $\ket{ij}\equiv\ket{i}\otimes\ket{j}$, with
$\sigma_z\ket{i}=(-1)^i\ket{i}$ and $i \in \{0,1\}$, 
and assuming an initial Bell state 
$\ket{\varphi}=\frac{1}{\sqrt{2}}(\ket{00}+\ket{11})$, we obtain
\begin{equation}
\rho(t) =\frac{1}{2} \left(
\begin{array}{cccc}
  1 & 0 & 0 & q(t) e^{i(\Omega_A+\Omega_B)t} \\
  0 & 0 & 0 & 0  \\
  0 & 0 & 0 & 0  \\
  q^*(t) e^{-i(\Omega_A+\Omega_B)t} & 0 & 0 & 1  \\
  \end{array} \right),
\label{eq:densityoperator}
\end{equation}
where the coherence decay factor $q(t)$ reads~\cite{paladino02,galperin06}
\begin{equation}
q(t)=e^{-\frac{ivt}{2}}\Big[A e^{-\frac{1}{2}\gamma(1-\alpha)t}+
 (1-A) e^{-\frac{1}{2}\gamma(1+\alpha)t}\Big],
\label{eq:coherence-decayfactor}
\end{equation}
with $\alpha=\sqrt{1-g^2}$, $A=\frac{1}{2}(1+\frac{1}{\alpha})$ and $g=v/\gamma$.
In the following we shall exploit $\rho(t)$ given by Eq.(\ref{eq:densityoperator}) to analyze 
the two-qubit entanglement dynamics.

\section{Entanglement dynamics\label{sec:entanglement-dynamics}}
To quantify the degree of entanglement of the system state $\rho(t)$ 
we use the entanglement of formation $E_f$~\cite{bennett96} that can be readily calculated by the formula \cite{wootters98}
\begin{equation}
E_f(\rho(t))=f\big(C(\rho(t))\big)=h\Big(\frac{1+\sqrt{1-C(\rho(t))^2}}{2}\Big), 
\label{eq:entanglement_formation}
\end{equation}
where $C(\rho(t))$ is the concurrence and $h(x)=-x \log_2 x -(1-x) \log_2 x(1-x)$. 
For the state $\rho(t)$ of Eq.~(\ref{eq:densityoperator}) we obtain $C(\rho(t))=|q(t)|$, where $q(t)$ is given in Eq.~(\ref{eq:coherence-decayfactor}). 
It is worth to notice that the evolved state $\rho(t)$  belongs to the Hilbert space
spanned by the  Bell states $\ket{\phi_\pm}=\frac{1}{\sqrt{2}}(\ket{00}\pm\ket{11})$, therefore the
entanglement of formation equals the entanglement 
cost~\cite{vidal2002PRL}.
In the  strong coupling regime, $g=v/\gamma>1$~\cite{paladino02,galperin06} entanglement 
revivals occur during the system dynamics~\cite{revivalstochastic,lofranco2012}.

\subsection{Dephasing under a "static" RT process}
To understand the nature of this phenomenon we initally consider the limiting case of an extraordinarily
slow RT process, $\gamma \to 0$ ($g \to \infty$).
This regime physically describes situations where the stochastic process is slow enough to be considered
static during the system time evolution lasting $t$, i. e. we assume $1/\gamma \gg t$~\cite{FalciPrl05}.
The evolution expressed by Eq.(\ref{eq:densityoperator}) describes an average 
resulting from the collection of several time evolutions each lasting $t$.
The average includes the possibility that the RT process takes any of the two values $\xi=0$ or 
$\xi=v$ at time $t=0$ with equal probability.
By a straightforward calculation we find that the concurrence in this case is 
given by $C\big(\rho(t)\big)=|\cos({vt}/{2})|$. Under these conditions we do not find any entanglement decay,
rather the concurrence is equal to one at times ${t}_n=2n\pi/v$ and it vanishes 
at times $\tilde{t}_n=(2n+1)\pi/v$, where $n$ is a non-negative integer.
\begin{figure}[t!]
\begin{center}
\includegraphics[width=0.65\textwidth]{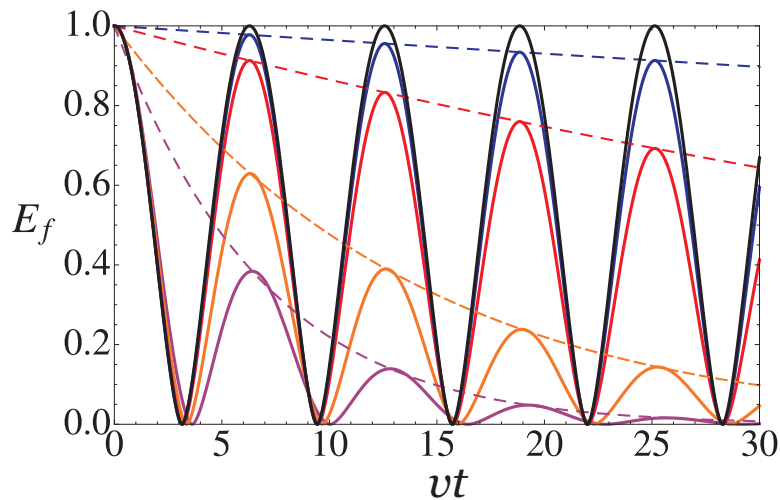}
\end{center}
\caption{\label{fig:RT}\footnotesize (color online) 
Entanglement of formation $E_f(\rho(t))$
as a function of the dimensionless time $vt$ 
for the 
$\ket{\varphi}=\frac{1}{\sqrt{2}}\big({\ket{00}+\ket{11}}\big)$.
From top to bottom,
the solid curves correspond to 
$g=\infty,200,50,10$ and $5$,
the dashed curves represent 
$f(e^{-\gamma t/2})$ for the same values of $\gamma=v/g$.}
\label{fig1}
\end{figure}
The entanglement  revivals  (see the top solid 
curve in Fig. \ref{fig:RT}) are not due to periodic entanglement death and rebirth 
by nonlocal operations. 
Indeed, the Hamiltonian evolution described by Eq.~ (\ref{eq:Hamiltonian1})  only includes local 
operations.
Since local operations cannot increase entanglement \cite{PlenioReview,HorodeckiReview}, its increase
during the intervals $]\tilde{t}_n,{t}_n]$ must be attributed 
to the manifestation of quantum correlations that were already present at 
times $\tilde{t}_n$, but were \emph{hidden}, in the sense 
that the density operator formalism does not capture them.

These correlations are evident in the quantum trajectory description of the system dynamics~\cite{carmichael}.
The system evolution in fact results from averaging on only two possible quantum trajectories. The first trajectory 
corresponds to $\xi(t)=0$ and the Bloch vector of qubit $A$ rotates around its $z$-axis with 
frequency $\Omega_A/(2\pi)$. The second trajectory orresponds to $\xi(t)=v$ and
the Bloch vector of qubit $A$ rotates around the $z$-axis with a different 
frequency $(\Omega_A+v)/(2\pi)$. The Bloch vector of qubit $B$ instead rotates in both cases around 
its $z$-axis with frequency $\Omega_B/(2\pi)$. Thus, during the first trajectory the system state evolves,
up to an  irrelevant global phase factor, as
\begin{equation}
\ket{\varphi_0(t)}=\frac{1}{\sqrt{2}}\Big(\ket{00}+
e^{-i(\Omega_A+\Omega_B)t}\ket{11}\Big) \, ,
\label{eq:state-1}
\end{equation}
while during the second trajectory the system evolves, apart 
from an irrelevant global phase factor, as
\begin{equation}
\ket{\varphi_v(t)}=\frac{1}{\sqrt{2}}\Big(\ket{00}+
e^{-ivt}e^{-i(\Omega_A+\Omega_B)t}\ket{11}\Big).
\label{eq:state-2}
\end{equation}
The two  quantum trajectories only differ by the fact that the basis states
$\ket{00}$ and $\ket{11}$ 
they acquire the additional relative phase ${ivt}$ in the quantum 
superpositions of Eqs.~(\ref{eq:state-1}) and (\ref{eq:state-2}). 
Since the two quantum trajectory occur with equal probability, 
the system's state is described by the quantum ensemble
\begin{equation}
{\cal A}=\Big\{\Big(p_0,\ket{\varphi_0(t)}\Big),\Big(p_v,\ket{\varphi_v(t)}\Big)\Big\},
\end{equation}
where $p_0=p_v=\frac{1}{2}$.
The entanglement associated to the quantum ensemble $\cal A$ can be suitably identified
by its average entanglement given by~\cite{bennett96,Cohen98,Nhal04,Carvalho07,footnote}
\begin{equation}
E_{av}\big({\cal A},t\big)=\sum_{i\in\{0,v\}} p_i E\big(\ket{\varphi_{i}(t)}\big)=1,
\label{eq:average_entanglement}
\end{equation}
since both states $\ket{\varphi_{0}(t)}$ and $\ket{\varphi_{v}(t)}$
are maximally entangled at any time ($E$ is the 
\textit{entropy of entanglement}~\cite{PlenioReview,HorodeckiReview}).

The \emph{hidden entanglement}~\cite{darrigo2012} of the ensemble $\cal A$ is defined 
as the difference 
between the average entanglement of Eq.~(\ref{eq:average_entanglement})
and the entanglement of the corresponding density operator 
$\rho(t)=\sum_i p_i \ketbra{\varphi_{i}(t)}{\varphi_{i}(t)}$:
\begin{equation}
E_h({\cal A},t)=E_{av}\big({\cal A},t\big)-E_f(\rho(t)).
\label{eq:hidden_entanglement}
\end{equation}
The meaning of hidden entanglement is simple: it measures the entanglement that cannot
be exploited as a resource due to the lack of classical knowledge 
about which state in the ensemble $\cal A$ we are dealing with. Once this 
classical information is provided, the entanglement can be recovered. 

Coming back to the interpretation of entanglement revivals, the ensemble description (average 
entanglement $E_{av}({\cal A},t)=1$)  tells us that at times $\tilde{t}_n$ the system is always 
in a maximally entangled state 
($\ket{\varphi_0(\tilde{t}_n)}$ or $\ket{\varphi_v(\tilde{t}_n)}$) but 
the lack of classical knowledge about which of the two states in the ensemble $\cal A$ we are dealing with
prevents us from distilling any entanglement: in fact, entanglement is {\em hidden} being
$E_f(\rho(\tilde{t}_n))=0$ and ${E_h}({\cal A}(\tilde{t}_n))=1$.
At times ${t}_n$ this lack of knowledge
is irrelevant since the random relative phase becomes meaningless at  
${v{t}_n}=2n\pi$: all the initial entanglement is recovered, 
$E_f(\rho({t}_n))=1$ and ${E_h}({\cal A}({t}_n))=0$.

\subsection{Dephasing due to a RT process: dynamic case}
We now investigate the case when the RT process evolves during the systen evolution
time, i. e. we consider the regime $1/\gamma \gtrsim t$. 
This situation is illustrated in Fig.~\ref{fig:RT} where we observe that the amplitude of revivals decreases as 
$\gamma$ increases (g decreases). 
Aslo in this case there is hidden entanglement. 
The possible quantum trajectories the system undergoes are now infinite.
The system state is described by the quantum ensemble 
${\cal A}(t)=\{P[\xi(t)],\ket{\varphi_\xi(t)}\}$ and the 
average entanglement of $\cal A$ is calculated by solving the path-integral
\begin{equation}
E_{av}({\cal A},t)=\int {\cal D}[\xi(t)] P[\xi(t)]\,E\big(\ket{\varphi_\xi(t)}\big).
\end{equation}
Once again, we obtain $E_{av}({\cal A},t)=1$ since during each trajectory
the state remains in a maximally entangled state at any time. 
On the other hand, the entanglement of formation assumes lower values with
respect to the static noise case, $\gamma \to 0$. In particular, the amplitude of revivals does
not reach anymore the initial maximum value. This is due to the fact that, in general,
the action of the RT process during the time evolution 
makes the coherences of $\ket{\varphi_\xi(t)} \bra{\varphi_\xi(t)}$  
(in the basis $\{|00\rangle,|11\rangle\}$) 
no longer in phase at  times ${t}_n$. 
In the time interval $]0,{t}_n]$ one or more 
transitions can occur between the two RT states, such that  
we can have a random extra phase at the times ${t}_n$ given by
\begin{equation} 
\vartheta({t}_n)\,=\,\int_0^{{t}_n} dt' \xi(t')- 2 \pi n,
\label{eq:vartheta}
\end{equation}
where $n$ is a non-negative integer.
This unknown phase difference is responsible for the decay of the absolute values of
coherences $|q({t}_n)|$ in the evolved two-qubit state $\rho({t}_n)$ of Eq.~(\ref{eq:densityoperator}):
if we knew the phase difference $\vartheta({t}_n)$ 
for each state $\ket{\varphi_\xi(t)}$,
we would be able to restore the coherence absolute value to 1,
and therefore recover all the initial entanglement, by simply applying the unitary local operation 
$e^{-i\frac{\vartheta({t}_n)}{2}\sigma_{z_A}}$.
For completeness we point out, that the relative maxima of
the entanglement of formation occur at $t_n^*=t_n/\sqrt{1-\frac{1}{g^2}}$, as one can
derive from equation (\ref{eq:coherence-decayfactor}).

Notice that the amount of the decay of the amplitude of entanglement revivals
is  monotonously related to the normalized 
autocorrelation function of the symmetric RT process 
$R(\tau)=\langle\xi(\tau)\xi(0)\rangle/\langle\xi^2(0)\rangle=e^{-\gamma t}$~\cite{Papoulis}. 
Indeed, as we have already mentioned, the reduction of the amplitude revivals is related
to the transitions of the RT occurred in $[0,{t}_n]$, whose 
mean number is $\gamma {t}_n/2$. 
From a quantitative point of view, for $g>1$ the coherences decay factor
Eq.~(\ref{eq:coherence-decayfactor}) can be approximated as 
$|{q}(t)|\simeq e^{-\gamma t/2}[\cos(vt/2)+1/g\sin(vt/2)]$, so that
$C(\rho({t}_n))\simeq e^{-\gamma {t}_n/2}$ and  $E_f(\rho({t}_n/2))=f(-\gamma {t}_n/2)$,
with $f$ defined in Eq.~(\ref{eq:entanglement_formation}).
This clearly shows that the decay of the entanglement amplitude revivals is due to the decrease of the 
RT correlations, or in other words, to the memory loss of the stochastic process $\xi(t)$
itself.

\section{Conclusions\label{conclusions}}
In this paper we have exploited the concept of hidden entanglement to interpret the occurrence of entanglement 
revivals in a particular system where back-action from the environment is absent. Namely, we have considered a 
system composed of two noninteracting qubits where one qubit is subject to random telegraph noise at pure 
dephasing. During the dynamics, entanglement vanishes and  revives always "remaining" inside
the system, as it is signalled by the average entanglement, $E_{av}(t)=1$ at any time.
At certain times $t_n$ this entanglement is completely hidden, in the sense that the entanglement of formation 
$E_f(t_n)=0$ while the hidden entanglement
$E_{h}(t_n)=E_{av}(t_n)-E_f(t_n)=1$. For this reason, the two-qubit entanglement can be simply recovered at 
subsequent times without the help of any nonlocal operation: in the considered case in fact the Hamiltonian 
only involves local operations.

Finally we remark that the concept of hidden entanglement can be of practical relevance in solid-state 
devices, where dominant noise sources tipically have large amplitude components at low frequencies. 
In these systems entanglement revivals may also be induced by applying 
local pulses to the qubits~\cite{lofranco2012,darrigo2012,falci04}.

%\ack
%This work was partially supported by EU through 
%Grant No. PITN-GA-2009-234970, and by the joint Italian-Japanese Laboratory on
%"Quantum Technologies: Information, Communication and Computation" of the 
%Italian Ministry of Foreign Affairs.


\section*{References} 

\begin{thebibliography}{50}
\bibitem{nielsen-chuang}
Nielsen M A and Chuang I L 2000
\textit{Quantum computation and quantum information}
(Cambridge University Press, Cambridge).

\bibitem{benenti-casati-strini}
Benenti G, Casati G and Strini G 2007
\textit{Principles of quantum computation and information}, vol. II
(World Scientific, Singapore).

\bibitem{PlenioReview}
Plenio M B and Virmani S 2007 {\it Quant. Inf. Comput.} \textbf{7} 1

\bibitem{HorodeckiReview}
Horodechi R \textit{et al} 2009 {\it Rev. Mod. Phys.} \textbf{81} 865

\bibitem{paladinoNJP}
Paladino E, D'Arrigo A, Mastellone A and Falci G 2011 
\textit{New J. Phys.} \textbf{13} 093037.

\bibitem{darrigoNJP}
D'Arrigo A and Paladino E 2012 \textit{New J. Phys.} \textbf{14} 05303.

\bibitem{bellomo2007PRL}
Bellomo B, Lo Franco R and Compagno G 2007 {\it Phys. Rev. Lett.} 
\textbf{99} 160502;
Bellomo B, Lo Franco R and Compagno G 2008 {\it Phys. Rev. A} 
\textbf{77} 032342.

\bibitem{lopez2008PRL}
L{\'{o}}pez C E, Romero G, Lastra F, Solano E and Retamal J C 2008 {\it 
Phys. Rev. Lett.} \textbf{101} 080503; L{\'{o}}pez C E, Romero G and 
Retamal J C 2010 {\it Phys. Rev. A} \textbf{81} 062114; Bai Y-K, Ye M-Y 
and Wang Z D 2009 {\it Phys. Rev. A} \textbf{80} 044301.

\bibitem{revivalstochastic}
Zhou D, Lang A and Joynt R 2010 {\it Quantum Inf. Process.} \textbf{9} 727;
Lo Franco R, D’Arrigo A, Falci G, Compagno G and Paladino E 2012
{\it Phys. Scripta} \textbf{T147} 014019.

\bibitem{rossini}
Rossini D, Benenti G and Casati G 2004 {\it Phys. Rev. A} \textbf{69} 
052317.

\bibitem{lofrancorandom}
Lo Franco R, Bellomo B, Andersson E and Compagno G 2012
{\it Phys. Rev. A} \textbf{85} 032318.

\bibitem{lofrancoreview}
Lo Franco R, Bellomo B, Maniscalco S and Compagno G 2012 {\em 
arXiv:1205.6419\/}

\bibitem{lofranco2012} Lo Franco R \textit{et al.} 2012 {\em in preparation}

\bibitem{darrigo2012} D'Arrigo A, Lo Franco R, Benenti G, Paladino E and 
Falci G 2012 {\em arXiv:1207.3294\/}

\bibitem{FalciPrl05}
Falci G, D'Arrigo A, Mastellone A and Paladino E 2005
{\it Phys. Rev. Lett.} \textbf{94} 167002.

\bibitem{solidstate}
Ithier G \textit{et al.} 2005 {\it Phys. Rev. B} \textbf{72} 134519;
Bylander J \textit{et al.} 2011 {\it Nat. Physics} \textbf{7} 565;
Chiarello F \textit{et al.} 2012 {\it New J. Phys.} \textbf{14} 023031.

%\bibitem{galperin06}
%Y. M. Galperin \textit{et al.}, %1,2,3,*, B. L. Altshuler4,5, J. 
%Bergli4, and D. V. Shantsev6,3
%Phys. Rev. Lett. \textbf{96}, 097009 (2006)
\bibitem{nakamura02}
Nakamura Y \textit{et al.} 2002 {\it Phys. Rev. Lett.} \textbf{88} 047901.

\bibitem{zorin96}
Zorin A \textit{et al.} 1996 {\it Phys. Rev. B} \textbf{53} 13682.

\bibitem{kafanov08}
Kafanov S \textit{et al.} 2008 % H. Brenning, T. Duty, and P. Delsing
{\it Phys. Rev. B} \textbf{78} 125411.

\bibitem{weissman}
Weissman M B 1988 {\it Rev. Mod. Phys.} \textbf{60} 537

\bibitem{paladino02}
Paladino E, Faoro L, Falci G and Fazio R 2002
{\it Phys. Rev. Lett.} \textbf{88} 228304.

\bibitem{paladino03}
Paladino E, Faoro L, D'Arrigo A and Falci G 2003
%Semiclassical analysis of 1/f noise in Josephson qubits, pubblicato 
%in \textit{Quantum Computing and Quantum Bits in Mesoscopic Systems}.
%, edito da A. Leggett, B. Ruggiero, P. Silvestrini, (2004) [ISBN 978-0-306-47904-5].
{\it Adv. Sol. St. Phys.} \textbf{43} 747.

\bibitem{galperin06}
Galperin Y M, Altshuler B L, Bergli J and Shantsev D V 2006
{\it Phys. Rev. Lett.} \textbf{96} 097009.

\bibitem{Papoulis}
Papoulis A 1965 \textit{Probability, Random Variables and Stochastic 
Processes}, New York, McGraw Hill.

\bibitem{PaladinoPhScr09}
Paladino E, D'Arrigo A, Mastellone A and Falci G 2009
{\it Phys. Scr.} \textbf{T137} 014017.

%\bibitem{PaladinoPRL02}
%Paladino E \textit{et al} 2002 {\it Phys. Rev. Lett.} \textbf{88} 228304.

\bibitem{bennett96}
Bennet C H \textit{et al} 1996 {\it Phys. Rev. A} \textbf{54} 3824.

\bibitem{wootters98}
Wootters W K 1998 {\it Phys. Rev. Lett.} \textbf{80} 2245.

\bibitem{vidal2002PRL}
Vidal G, D{\"{u}}r W, and Cirac J I,
{\it Phys. Rev. Lett.} 2002 \textbf{89} 027901.

%\bibitem{paladino05}
%Paladino E, Mastellone A, D'Arrigo A and Falci G 2005
%in \textit{Realizing Controllable Quantum States}
%: Proceedings of the Mesoscopic Superconductivity and Spintronics 
%(Ms+S2004), edito da H. Takanayanagi e J. Nitta,
%(World Scientific Publishing Company) % [ISBN 981-256-468-3].

\bibitem{carmichael}
Carmichael H J 1993 {\em An Open Systems Approach to
Quantum Optics}, Springer, Berlin.

\bibitem{Cohen98}
Cohen O 1998 {\it Phys. Rev. Lett.} \textbf{80} 2493.

\bibitem{Nhal04}
Nha H and Carmichael H J 2004
{\it Phys Rev. Lett.} \textbf{93} 120408.

\bibitem{Carvalho07}
Carvalho A R R, Busse M, Brodier O, Viviescas C and Buchleitner A 2007
{\it Phys. Rev. Lett.} \textbf{98} 190501.

\bibitem{footnote}
Note that in Ref.~\cite{Cohen98} the expression
``hidden entanglement'' is used with a different meaning.

\bibitem{falci04}
Falci G, D’Arrigo A, Mastellone A, and Paladino E
2004 {\it Phys. Rev. A} {\bf 70} 40101(R).

%\bibitem{Slichter}
%Slichter C P 1990 \textit{Principles of Magnetic Resonance}, Springer.

%\bibitem{mukhtar2010PRA1}
%Mukhtar M, Saw T B, Soh W T, and Gong J 2010
%{\it Phys. Rev. A} \textbf{81} 012331;
%Mukhtar M, Soh W T, Saw T B, and Gong J 2010
%\textit{ibid.} \textbf{82}, 052338.

\end{thebibliography}
\end{document}